\documentclass[a4paper]{VisionStyle}
\usepackage{epsfig}

\begin{document}

\title{A Her X-1 Turn-On: Using the pulse profile to probe the outer edge
  of an accretion disk}

\author{M.\,Kuster\inst{1} \and J.\,Wilms\inst{1} \and R.\,Staubert\inst{1}
  \and P.\,Risse\inst{1} \and W.A.\,Heindl\inst{2} \and R.\,Rothschild\inst{2} \and
  N.I.\,Shakura\inst{3} \and K.A.\,Postnov\inst{3,4}}

\institute{Institut f\"ur Astronomie und Astrophysik, Abt. Astronomie,
  Sand 1, D-72076 T\"ubingen, University of T\"ubingen,
  Germany 
  \and
  Center for Astrophysics and Space Sciences, UCSD, La
  Jolla, CA 92093, USA 
  \and
  Sternberg Astronomical Institute, Moscow University, 119899 Moscow,
  Russia 
  \and 
  Faculty of Physics, Moscow State University, 119899 Moscow, Russia
  }

\maketitle 

\begin{abstract}
  The X-ray binary pulsar Her X-1 shows a wide variety of long and short
  term variabilities in the X-ray light curve. The 35\,d variability of the
  source is interpreted as the influence of a warped, inclined, and twisted
  accretion disk periodically covering the line of sight to the central
  neutron star. In 1997 September we observed the ``turn-on'' of a 35\,d
  cycle with the Rossi X-ray Timing Explorer ({\sl RXTE}). Spectral
  analysis reveals that during early phases of the turn-on the overall
  spectrum is composed of X-rays scattered into the line of sight plus
  heavily absorbed X-rays. This interpretation is consistent with the
  variation of the pulse profile observed at the same time. The overall
  shape of the pulse profile is not changing, but towards earlier phases of
  the turn-on the pulse signature is steadily ``washed out''. This behavior
  can be understood as an influence of scattering and absorption due to the
  presence of the accretion disk rim. Using a Monte Carlo code we simulate
  the influence of both processes on a time variable, beamed emission
  characteristic, similar to the pulse profile of Her X-1. By comparing the
  results of the simulation with the observed profiles we determine the
  amount of scattered radiation, absorbed radiation, and the size of the
  scattering region.
  
\end{abstract}

\section{Introduction}\label{sect:intro}
Her X-1/HZ Her is a X-ray binary system consisting of a neutron star Her
X-1 and its optical companion HZ Her. The system rotates with a 1.7\,d
orbital period around their common center of mass. The X-ray light curve of
Her X-1 shows strong variability on many time scales: e.g. 1.24\,s
pulsations originating in the rotational period of the neutron star, or the
35\,day intensity variation caused by a precessing, inclined, warped, and
twisted accretion disk
(see~\cite{mkuster-C1:petterson75a,mkuster-C1:schandl94a}).

\begin{figure}
  \begin{center}
    \includegraphics[width=0.97\columnwidth]{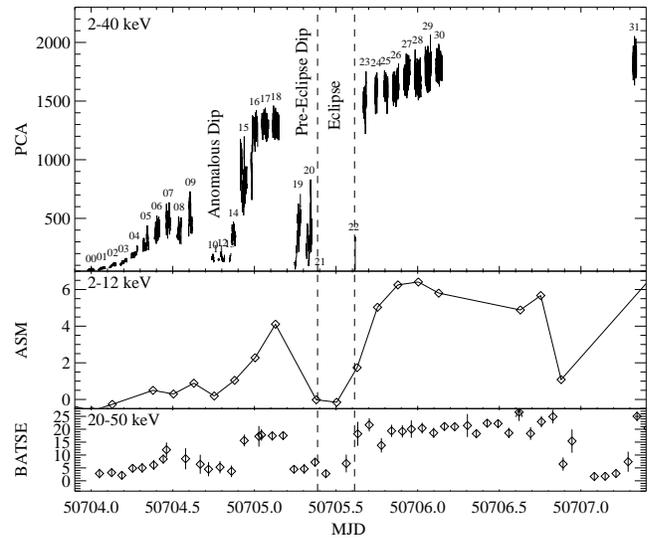}
  \end{center}
  \caption{From top to bottom: {\sl RXTE PCA}, {\sl RXTE ASM}, and {\sl CGRO BATSE} light
    curve over the time of the turn-on. The X-ray flux is steadily
    increasing from observation~00 until observation~31. During the time of
    the turn-on an eclipse, anomalous and pre-eclipse dip occurred. For a
    further description see text.}
  \label{mkuster-C1_fig:fig1}
\end{figure}

During the 35 day cycle the X-ray intensity shows strong variation, with
two maxima in intensity called the ``main-on'' and the ``short-on''. This
intensity variation is caused by periodic covering of the neutron star by
the outer (main-on) or the inner edge (short-on) of the accretion disk. The
``turn-on'' marks the beginning of the main-on and therefore the beginning
of the 35 day cycle. During the turn-on the outer edge of the accretion
disk steadily opens the line of sight to the neutron star. While the outer
accretion disk rim moves out of the line of sight, the observed X-ray
intensity steadily increases, until it reaches the maximum after about 2-3
days. Fig.~\ref{mkuster-C1_fig:fig1} shows the X-ray light curve of a
complete turn-on observed in 1997 September with {\sl RXTE PCA}.  In this
paper we present preliminary results of our combined spectral and temporal
analysis, and resulting simulated pulse variation over the turn-on.

\section{Turn-on light curve and spectral analysis}
\label{mkuster-C1_sec_spectral}
The complete X-ray light curve observed with {\sl RXTE} is shown in
Fig.~\ref{mkuster-C1_fig:fig1}. During the turn-on the observed flux
steadily increases until its maximum in orbit~31. Around MJD 50704.8 and
MJD 50705.3 an anomalous dip, a pre-eclipse dip, and at MJD 50705.5 an
eclipse occurred. For our spectral modeling and pulse analysis we ignored
the times of the dips (orbit.~10--14) and the time of the eclipse
(orbit~19--22) because our spectral model is not applicable during these
times. The numbers in the top panel of Fig.~\ref{mkuster-C1_fig:fig1}
identify individual {\sl RXTE} orbits starting from~00 to~31.

\begin{figure}
  \begin{center}
    \includegraphics[width=0.97\columnwidth]{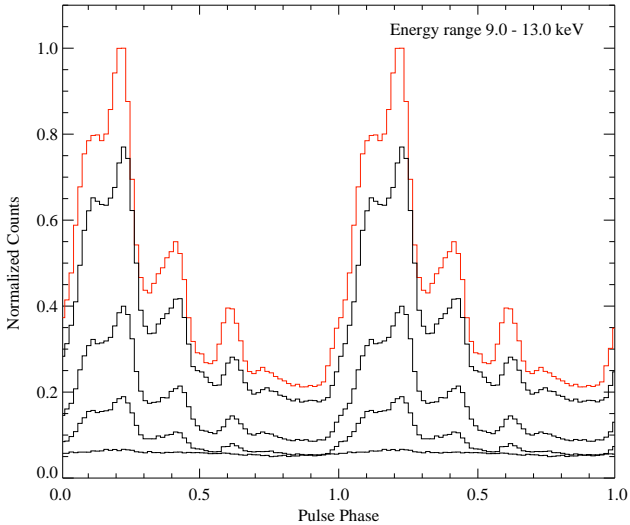}
  \end{center}
  \caption{Evolution of the pulse profile in the energy range from
    9.0 to 13.0 keV over the time of the turn-on. From top to bottom: Pulse
    profile of orbit~30, 23, 20, 17, and 04. The pulse profile of orbit~30
    (red) is used as a template for the pulse simulations. All profiles are
    normalized relative to the maximum flux in the profile of orbit~30,
    which is set to unity. Pulse phase zero is defined as the maximum flux
    in the pulse profile, thus absolute phase information is lost.}
  \label{mkuster-C1_fig2}
\end{figure}

As presented earlier (e.g., \cite{mkuster-C1:kuster99a}), a spectral
analysis and an analysis of the pulse profile reveals that,
\begin{itemize}
\item the spectral evolution during the turn-on can be described by a
  combination of direct, absorbed, and scattered radiation,
\item the intrinsic shape of the pulse profile is not changing over the
  time of the turn-on, but the pulse signature is steadily washed out
  towards the beginning of the turn-on (see Fig.~\ref{mkuster-C1_fig2}),
  and
\item the pulse profile and the spectrum at the end of the turn-on (later
  than orbit~30) shows no significant signature of absorbed or scattered
  radiation.
\end{itemize}
These results are in agreement with earlier observations of e.g.
\cite*{mkuster-C1:davison77a}, \cite*{mkuster-C1:becker77a}, or
\cite*{mkuster-C1:parmar80a}, who found the contribution of two components
to the overall observed X-ray spectrum: a ``scattered'' component and a
``absorbed'' component. The origin for the scattered component is generally
interpreted (e.g.  \cite{mkuster-C1:mihara91a}) as the influence of an
extended, ionized accretion disk corona at a distance of $r < 5 \times
10^{11}$cm. Recent observations with {\sl XMM-Newton's RGS} show strong
indication for the presence of an accretion disk corona in Her X-1 as well
(\cite{mkuster-C1:burwitz02a}). The ``absorbed'' spectral component is
attributed to the influence of the outer accretion disk rim partially
obscuring the central emission region.

\section{Simulating pulse variation}
\begin{figure}
  \begin{center}
    \includegraphics[width=0.97\columnwidth]{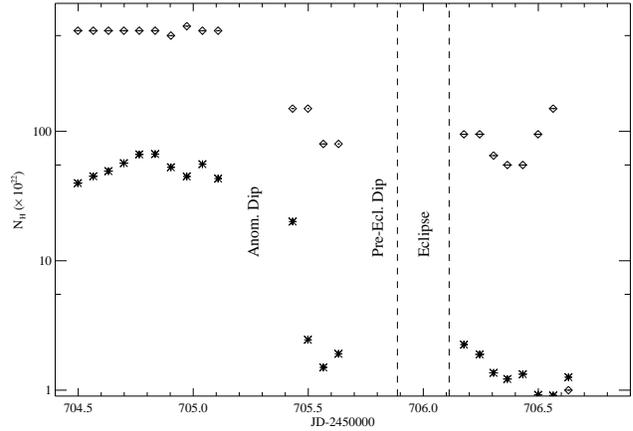}
  \end{center}
  \caption{$N_{\rm e,abs}$ and $N_{\rm e,ion}$ over the time of the
    turn-on. The astericks mark $N_{\rm abs,e}$ determined via the spectral
    analysis using an absorption model. The diamonds mark $N_{\rm e,ion}$
    we get from our pulse analysis. This $N_{\rm e,ion}$ can be attributed
    to a scattering medium which modifies the shape of the pulse profile.}
  \label{mkuster-C1_scattnh}
\end{figure}
\begin{figure*}
  \begin{center}
    \includegraphics[width=0.99\textwidth]{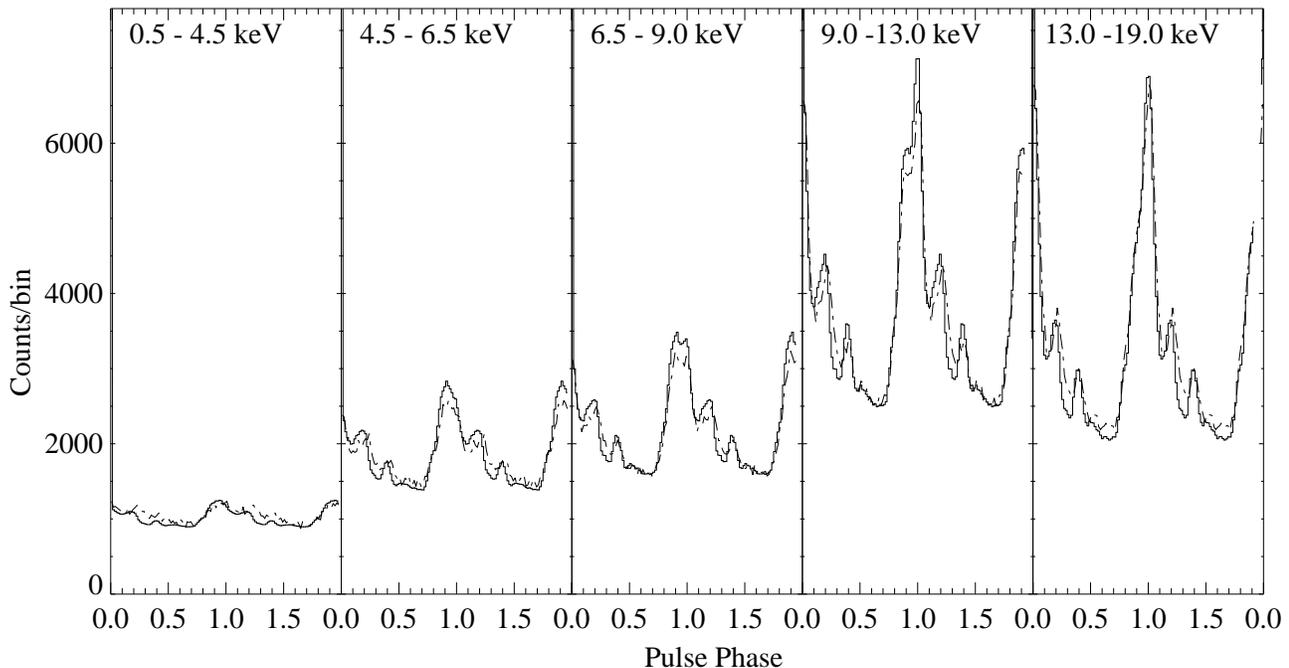}
  \end{center}
  \caption{Fit results for orbit~08. The dashed line marks
    the observed pulse profile and the solid line the simulated pulse
    profile. The simulated pulse profile is a modified version of the pulse
    profile of orbit~30 (see Fig.~\ref{mkuster-C1_fig2}).}
  \label{mkuster-C1_fit-res}
\end{figure*}
Taking the results from the spectral analysis into account, we can assume
that the variation of the pulse profile at the beginning of the turn-on is
caused by scattering and photoelectric absorption in the medium covering
the line of sight to the neutron star (the outer accretion disk rim and
corona). Furthermore we can assume that at the end of the turn-on, when the
accretion disk rim has moved out of the line of sight, the pulse profile
represents an unaffected emission characteristic. Thus we can use the pulse
profile observed in, e.g., orbit~30 as a template and investigate the
effects of a scattering and absorbing corona on the pulse shape depending
on $N_{\rm H}=N_{\rm e}$ and the size of the scattering region. 

For our simulations we used a Monte Carlo code developed at our Institute,
assuming simplified slab geometry and taking into account Compton
scattering, photoelectric absorption, and fluorescent line emission. As a
result we obtain the \emph{Green's function $G(t,t_{\rm 0})$} as a function
of \emph{$N_{\rm H}$=$N_{\rm e}$}, the \emph{thickness of the slab $d$},
and the \emph{energy band}, and angle-dependent photon spectra of photons
leaving the slab. We simulated $G(t,t_{\rm 0})$ for two cases, a fully
ionized corona and a neutral corona. The total Green's function for a
partially ionized corona can then be calculated as $G_{\rm tot}(t,t_{\rm
  0})=(1-a)\,G_{\rm ion}+ a\,G_{\rm abs}$, where a is the fraction of
ionization. For $G_{\rm abs}$ we used $N_{\rm e,abs}$ determined by the
spectral analysis. With a variable $N_{\rm e,ion}$ for $G_{\rm ion}$ we can
then use this simulated Green's function to calculate modified pulse
profiles an observer located at infinity would see according to:
\begin{equation}\label{eqn:green}
I^{\infty}(t)=\int_{-\infty}^{t} G(t,t_0)I(t_0)dt_0
\end{equation}
where $I(t_{\rm 0})$ is the intensity of the template pulse profile at time
$t_{\rm 0}$. Finally we perform a $\chi^{\rm 2}$ minimization fit of
\emph{simulated} to \emph{observed} pulse profiles for each single {\sl
  RXTE} orbit shown in Fig~\ref{mkuster-C1_fig:fig1}. As an example, the
result of the fit for orbit~08 is given in Fig.~\ref{mkuster-C1_fit-res}.
All four energy bands were fitted simultaneously. By combining this
procedure with a spectral analysis based on a partial covering model (see
\cite{mkuster-C1:kuster99a}), we are able to separate the amount of
scattered and absorbed radiation from purely scattered radiation, and
determine the thickness $d$ of the corona as well as the uncertainties,
respectively.

\section{Results}
Preliminary results of our pulse analysis are shown in
Fig.~\ref{mkuster-C1_scattnh}. The crosses in Fig.~\ref{mkuster-C1_scattnh}
show the evolution of neutral density $N_{\rm H}=N_{\rm e,abs}$ deduced
from the spectral analysis. As reported earlier by us
(\cite{mkuster-C1:kuster01a}), $N_{\rm e,abs}$ is increasing at the
beginning of the turn-on and starts to decline orbit~05. This behavior can
be understood due to an additional flux contribution of a scattering corona
which comes important at high densities and thus dominates the spectrum
during early phases of the turn-on. The diamonds in
Fig.~\ref{mkuster-C1_scattnh} show the results we get from our pulse
profile analysis. This $N_{\rm e,ion}$ can be attributed to the flux
contribution of a scattering corona moving slowly out of the line of sight
to the neutron star.

\begin{acknowledgements}
This work is partially funded from DLR grant 50 0x0002 and NATO grant
PST CLG975254.
\end{acknowledgements}

\appendix

\end{document}